# Attacking an OT-Based Blind Signature Scheme

Stylianos Basagiannis, Panagiotis Katsaros and Andrew Pombortsis

*Abstract*—In this paper, we describe an attack against one of the Oblivious-Transfer-based blind signatures scheme, proposed in [1]. An attacker with a primitive capability of producing specific-range random numbers, while exhibiting a partial MITM behavior, is able to corrupt the communication between the protocol participants. The attack is quite efficient as it leads to a protocol communication corruption and has a sound-minimal computational cost. We propose a solution to fix the security flaw.

*Index Terms*—Oblivious Transfer, blind signatures.

## I. INTRODUCTION

OBLIVIOUS TRANSFER (OT) constitutes a powerful tool used today in modern cryptography. In the first introduction of the $OT_1^2$ mechanism by Rabin [2], it is assumed that in a communication system, Alice transmit to Bob a two-part message, where only the one part is the secret that Alice wants to share. From Bob's side, Bob does not know which one of the two is the real secret, so he selects one of them with probability ½ of success (or ½ of failure).

In the related bibliography, OT based security protocols [3] [4] that aim to guarantee a variety of security properties such as anonymity or privacy of the participants, especially when OT is combined with other cryptographic primitives, e.g. blind signatures [5]. Through these works, OT has been involved into several improvements regarding efficiency and of the OT-based communication systems. The basic $OT_1^2$, described above, has been replaced by mechanisms of $OT_1^n$ shown in [3], where the *Sender* dispatch *N* message to the *Chooser*, and the *Chooser* selects the appropriate message without knowing the initial selection of the *Sender*.

In reports [6] and [7], the OT mechanism is combined with various cryptographic techniques, in order to provide the involved participants with even more security guarantees. OT is combined with signature schemes providing strong fairness, anonymity and privacy of the communication. Both of the reports also provide a detailed analysis of the protocol in terms of anonymity and privacy. There are also reports like [8], where security threats over the OT-based protocol schemes have been classified into high and low cost attacks. All these kinds of security threats are managed to succeed regarding the computational cost of the encryption used between the protocols' participants, where an adversary is consider containing the maximum computational power, performing a variety of attack actions.

In this paper, we present an attack against the OT-based double blind signature scheme protocol described in [1]. More precisely, we show that a partial MITM intruder with a low computational cost can corrupt the protocol's communication by integrity violating one of the protocols' exchanged messages (by tagging specific random numbers). As a result the protocols' agents will accept the corruption occurred, which misinterprets the overall communication.

## II. THE OT-BASED BLIND SIGNATURE PROTOCOL

This paper focuses on the analysis of a variant *1*-out-of-*n* Oblivious Transfer ($OT_1^n$) based on blind signatures protocol. The specific protocol incorporates a blind mechanism from both the *Sender*'s and the *Chooser*'s side. To achieve cryptographic efficiency the protocol involves a series of security perspectives such as public key cryptography, blind signatures and a keyed hash function. A random number generator for both of the participants is also used in order to overcome predictability attacks caused by an *Intruder*. The following notation is used throughout the paper:

| | |
|---|---|
| $N$ | RSA modulus |
| $\{S_0,...,S_{n-1}\}$ | *Sender* posses initially *n* secret strings $S_i$ |
| $\sigma$ | *Chooser* posses initially an integer $\sigma \in [0..n-1]$ |
| $H$ | Pre-agreed Hash Function |
| $SK<N,d>$ | Secret Key |
| $PK<N,e>$ | Public Key |
| $\{U_0,...,U_{n-1}\}$ | Random numbers $U_0,...,U_{n-1} \in Z_N^*$ |
| $SP(\ )$ | Secure padding scheme for RSA |
| $C$ | Random number $C \in Z_N^*$ |
| $R$ | Random number $R \in Z_N^*$ |
| $Y_\sigma$ | Blind Signature for the *Sender* |
| $K_{j=0..n-1}$ | Encryption Keys for the *Sender* |
| $\oplus$ | XOR operator |
| $d1>>d2$ | Right shift-bit operator |

The *Sender* has *n* input secret strings, $\{S_0,...,S_{n-1}\}$ and the *Chooser*'s input is an integer $\sigma \in [0..n-1]$. Because of the $OT_1^n$, the *Chooser* should learn a secret $S_\sigma$ and nothing on any other $\{S_0,...,S_{n-1}\} - \{S_\sigma\}$. On the other hand, the *Sender* should learn nothing about $\sigma$. The protocol described here provides unconditional protection for the *Chooser* and computational protection in the random oracle model for the *Sender*. Due to the OT operability, the specific protocol may be often invoked multiple times between the participants.

The basic steps of the described protocol are illustrated in Fig.1 and can be summarized as follows:

The authors are with the Dept. of Informatics, Aristotle University of Thessaloniki, 54124, Greece. Email:{basags, katsaros, apombo}@csd.auth.gr.

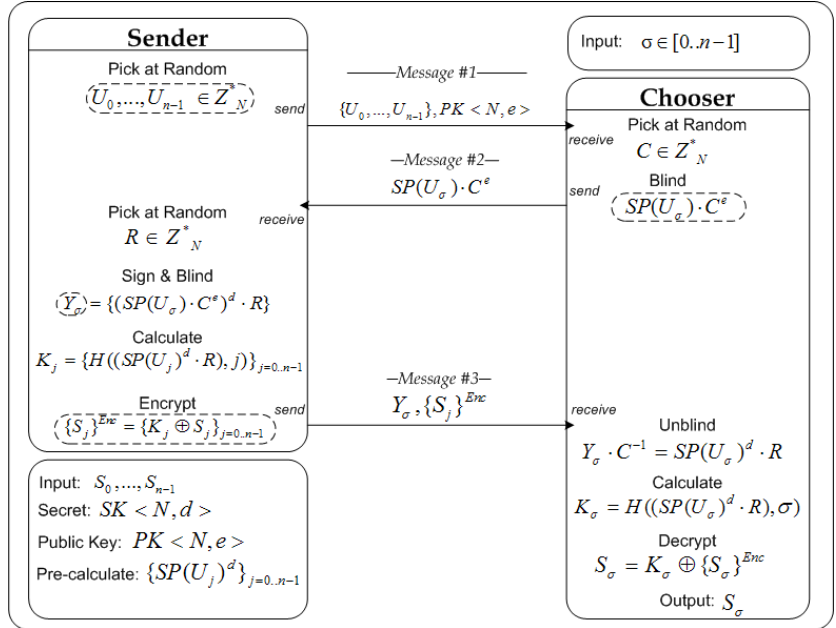

Fig. 1. The Double Blind Protocol Scheme Based on Oblivious Transfer

1) Initially the *Sender* picks $n$ random numbers $\{U_0,...,U_{n-1}\}$ and the *Chooser* inputs an integer $\sigma$ over the range $[0..n-1]$. Then, *Sender* selects his RSA keys and sends them along with the set of its numbers to the *Chooser*.

2) Upon the first message is receipted, the *Chooser* will randomly (according to $\sigma$) select a number $U_\sigma$ and compute the $SP(U_\sigma)$, which in turn, it is blinded with a random selected number $C$, in order to produce $SP(U_\sigma) \cdot C^e$. He then sends the message to *Sender*.

3) When the *Sender* receives the second message, he randomly selects his blind factor $R$ (random number). Then he signs (with $SK<N, d>$) and blinds (with $R$) the message, creating $Y_\sigma = \{(SP(U_\sigma) \cdot C^e)^d \cdot R\}$. On the next step, the *Sender* creates $K_j$ keys with $j=0..n-1$ using the pre-agreed hash function $H$. Those keys will be used to encrypt all the n string secret messages $\{S_0,...,S_{n-1}\}$ using the $\oplus$ operator.

4) Once the *Sender* completes the encryption, of $\{S_j\}^{Enc} = \{K_j \oplus S_j\}_{j=0..n-1}$, he dispatches the third message to the *Chooser*, which contains the blind signature $Y_\sigma$ and the encrypted set of messages $\{S_j\}^{Enc}$. At this point *Sender* accomplishes his participation to the protocol.

5) *Chooser* receives the third message and moves on to un-blinding $Y_\sigma$ with $C^{-1}$. While acquiring $Y_\sigma \cdot C^{-1} = SP(U_\sigma)^d \cdot R$, he uses his initial choice $\sigma$ to compute the appropriate $K_\sigma$ for revealing only the $S_\sigma$ secret message. $K_\sigma$ will be computed by applying the hash function $H$ to the remaining $SP(U_\sigma)^d \cdot R$, resulting to $K_\sigma = H((SP(U_\sigma)^d \cdot R), \sigma)$.

6) Once $K_\sigma$ has been computed, the *Chooser* will apply it into the encrypted set of messages $\{S_j\}^{Enc}$ obtaining at a final stage, $S_\sigma$ by $S_\sigma = K_\sigma \oplus \{S_\sigma\}^{Enc}$, and thus completing the protocol. At this point, the *Chooser* will have knowledge of $S_\sigma$ without being able to know anything else of the set $\{S_j\}^{Enc}$. As a security-consequence of the protocol's completion, the outcome will be the preservation of information-theoretic privacy for the *Sender* and computational privacy for the *Chooser* according to [1].

III. ANALYSIS OF THE ATTACK

Similar security analysis reports, such as [8], have indicated security threats that may be launched in OT-based communication systems. The majority of them are based over an intruder that eavesdrop all the protocols' messages, containing also a degree of computational power that may reveal to him, previously unknown information.

An *Intruder* may launch a series of actions that could lead the protocols' participants into security failures, such as a DoS attack from the *Chooser*'s side, or impersonations using previously recorded *Sender*'s messages in other (same protocol) sessions. While the protocol avoids those kinds of attacks by using double-blind signatures for both of its participants, it may allow *Intruder*'s integrity violations, by leading to a corrupt communication. In the rest of this section, we present the attack where an *Intruder*, being placed as a MITM entity may corrupt the proposed protocol.

*A. Description of the Attack*

The necessary prerequisites for the *Intruder* are the following:
a) The *Intruder* is a MITM entity among the protocols' participants that eavesdrops all the messages, b) he has the ability of generating random numbers over a specific range of values and c) he can concatenate previously intercepted messages with data that he has created. Fig. 2 provides a detailed description of the attack mounted against the protocol.

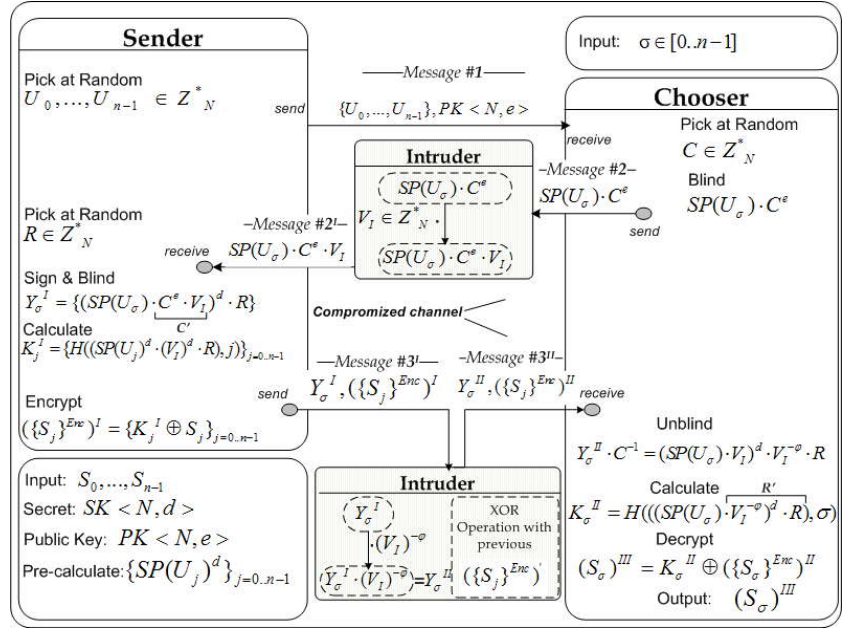

Fig. 2. The proposed attack

The proposed attack is described as follows:

1) In the first step, the *Intruder* leaves the first message uninterrupted to reach its intended destination. *Chooser* upon receipt of the message, selects, under uniform probability distribution, a number $U_\sigma$ according to his input σ that he had initially chosen; then he generates random number $C$ blinding $SP(U_\sigma)$ to form the second message.

2) Through a compromised channel, the *Intruder* receives the second message $SP(U_\sigma) \cdot C^e$ originated from the *Chooser*. Then, the *Intruder*, using a random number generator, corrupts the message with a random number $V_I \in Z^*$ creating $SP(U_\sigma) \cdot C^e \cdot V_I \xrightarrow{C^e \cdot V_I \equiv C^I} SP(U_\sigma) \cdot C^I$. The *Sender* will then create $(SP(U_\sigma) \cdot C^I)^d$, blinding it with $R \in Z^*_N$, having $(C^I \cdot R) \in Z^*_N \Rightarrow (V_I \cdot C) \in Z^*_N$, with the overall signature being $Y_\sigma^I = \{(SP(U_\sigma) \cdot C^I)^d \cdot R\}$.

3) Producing the appropriate keys $K_{j=0..n-1}$, the corrupted product will be intersected into the $H((SP(U_j)^d \cdot (V_I)^d \cdot R)$ creating falsie $K_j^I$ s. As a consequence, the encrypted set $\{S_j\}^{Enc}$ will be encrypted using the wrong keyset.

4) *Sender* dispatches message $3^I$ to the *Intruder* who from his side, computes again a random $(V_I)^{-\varphi}, \varphi \in \aleph$ down-blinding signature $Y_\sigma^I$ to $Y_\sigma^I \xrightarrow{\cdot(V_I)^{-\varphi}} Y_\sigma^{II}$ with $C \cdot V_I^{d-\varphi}$. Additionally, the *Intruder* sets $\{\{S_j\}^{Enc}\}^I \oplus \{\{S_j\}^{Enc}\}' = \{\{S_j\}^{Enc}\}^{II}$ using a previously intercepted set of $\{\{S_j\}^{Enc}\}'$, forwarding the new message $3^{II}$ to the *Chooser*.

5) *Chooser* un-blinds signature $Y_\sigma^{II}$ with $C^{-1}$ in order to calculate from his side the right key $K_\sigma$ (false $K_\sigma^I$) to decrypt his initial choice $S_\sigma$. Due to the previous *Intruder's* operation, *Chooser* manages to decrypt the wrong $S_\sigma$ ($(S_\sigma)^{III}$) without knowing it, and thus accepting the corruption made.

IV. FIXING THE FLAW AND CONCLUSION

When exchanging the second and the third message, the participants may re-use the already implemented hash function $H$ which is predetermined by the protocols' specifications upon the agents. While sealing $C^e$ with the $f_\sigma = (C^e, \sigma)$ where $f_\sigma$ is the keyed hash function of $H$ from the *Chooser's* side, and $f_R = (Y_\sigma, R)$ from the *Sender's* side, any alteration made to the exchanged messages will be discarded. Such an optimization requires one more of execution of the keyed hash function for both of the participants when the protocol is completed, in order to identify possible intruder corruptions.

The presented OT-based blind signature schemes have been adopted in systems such as e-auctions and online-gaming, where privacy remains the first objective. However, such communication prototypes have to provide, at the end, accuracy of the delivered services, when the protocol finalizes.